\documentclass[11pt]{article}
\usepackage[utf8]{inputenc}
\usepackage{amsmath}
\usepackage{amssymb}
\usepackage{tensor}
\usepackage{mathrsfs}
\usepackage{cite}
\usepackage{multirow}
\usepackage{framed}
\usepackage{graphicx}
\usepackage[colorlinks=true,urlcolor=blue,anchorcolor=blue,citecolor=blue,filecolor=blue,linkcolor=blue,menucolor=blue,linktocpage=true,pdfproducer=medialab,pdfa=true]{hyperref}
\usepackage{calc}
\usepackage[symbol]{footmisc}
\usepackage{color}
\usepackage{url}
\usepackage{float}

\newcommand\bom[1]     {{\mbox{\boldmath $#1$}}}

\newcommand\CT {\text{CT}}

\makeatletter

\@addtoreset{equation}{section}

\def\draftdate{\relax}
\def\mda{\relax}
\def\mua{\relax}
\def\mla{\relax}
\def\draft{
\def\thtystars{******************************}
\def\sixtystars{\thtystars\thtystasr}
\typeout{}
\typeout{\sixtystars**}
\typeout{* Draft mode!
         For final version remove \protect\draft\space in source file *}
\typeout{\sixtystars**}
\typeout{}
\def\draftdate{\today}
\def\mua{\marginpar[\boldmath\hfil$\uparrow$]%
                   {\boldmath$\uparrow$\hfil}%
                    \typeout{marginpar: $\uparrow$}\ignorespaces}
\def\mda{\marginpar[\boldmath\hfil$\downarrow$]%
                   {\boldmath$\downarrow$\hfil}%
                    \typeout{marginpar: $\downarrow$}\ignorespaces}
\def\mla{\marginpar[\boldmath\hfil$\rightarrow$]%
                   {\boldmath$\leftarrow $\hfil}%
                    \typeout{marginpar: $\leftrightarrow$}\ignorespaces}
\overfullrule 5pt
\oddsidemargin -12mm
\marginparwidth 29mm
}

\def\starline{\hfil\strut\hfil\hbox to \textwidth {\stasr}\hfil}

\def\beq{\begin{equation}}
\def\eeq{\end{equation}}
\def\bsp#1\esp{\begin{split}#1\end{split}}
\def\bal#1\eal{\begin{align}#1\end{align}}

\def\beeq{\begin{eqnarray}}
\def\eeeq{\end{eqnarray}}
\newcommand{\eps}      {\varepsilon}

\newcommand{\rd}       {{\mathrm{d}}}

\newcommand\tsig[1]    {\sigma^{\mathrm{#1}}}
\newcommand\tsigk[4]    {\sigma^{{\rm #1}^{#2} {\rm #3}^{#4}}}
\newcommand\dsig[1]    {\rd\sigma^{{\rm #1}}}
\newcommand\dsigk[4]    {\rd\sigma^{{\rm #1}_{#2} {\rm #3}_{#4}}}
\newcommand\dsiga[2]   {\rd\sigma^{{\rm #1,A}_{\scriptscriptstyle #2}}}

\newcommand{\cII}[1] {{\cal I}\kern-4pt *\kern-4pt{\cal I}_{#1}}
\newcommand{\cIJ}     {{\cal I}\kern-4pt *\kern-4pt{\cal J}}
\newcommand{\cJJ}[1] {{\cal J}\kern-4pt *\kern-4pt{\cal J}_{#1}}
\newcommand{\cJI}     {{\cal J}\kern-4pt *\kern-4pt{\cal I}}
\newcommand{\cJK}     {{\cal J}\kern-4pt *\kern-4pt{\cal K}}
\newcommand{\cKJ}[1] {{\cal K}\kern-4pt *\kern-4pt{\cal J}_{#1}}
\newcommand{\cKI}     {{\cal K}\kern-4pt *\kern-4pt{\cal I}}
\newcommand{\cKK}     {{\cal K}\kern-4pt *\kern-4pt{\cal K}}

\newcommand{\bSCS}[1]  {\bom{\mathrm C}\kern-2pt\bom{\mathrm S}_{#1}}

\newcommand{\cSCS}[2]  {{\cal C}\kern-2pt{\cal S}_{#1}^{#2}}

\newcommand{\IcSCS}[2]  {\mathrm{C}\kern-2pt\mathrm{S}_{#1}^{#2}}
\def\s12{s_{12}}

\newcommand{\colorful}{CoLoRFulNNLO }
\newcommand{\nnlocal}{{\tt NNLOCAL}}

\begin{document}

\numberwithin{equation}{section}

\begin{titlepage}
\noindent
 BONN-TH-2026-12\hfill May 2026\\
\vspace{0.6cm}
\begin{center}
{\LARGE \bf 
    \colorful for color-singlet processes:\\ An update on \nnlocal\footnote{Presented at Loops and Legs in Quantum Field Theory (LL2026), 12-17 April 2026, Bayreuth (Germany).}
}
\vspace{1.0cm}

\large
S.~Van Thurenhout$^{\, a}$\footnote[2]{Speaker}, V. Del Duca$^{\,b}$, C. Duhr$^{\, c,d}$, L. Fek\'esh\'azy$^{\,e,f}$, F. Guadagni$^{\, g}$, P. Mukherjee$^{\, e}$, G. Somogyi$^{\, a}$ and F. Tramontano$^{\, h}$\\
\vspace{0,5cm}
\normalsize
{\it $^{\, a}$HUN-REN Wigner Research Centre for Physics, Konkoly-Thege Mikl\'os u. 29-33, 1121 Budapest, Hungary}\\
{\it $^{\, b}$INFN, Laboratori Nazionali di Frascati, 00044 Frascati (RM), Italy}\\
{\it $^{\, c}$Bethe Center for Theoretical Physics, Universität Bonn, D-53115, Germany}\\
{\it $^{\, d}$Cluster of Excellence “Color meets Flavor”, Universität Bonn,
D-53115 Bonn, Germany}\\
{\it $^{\, e}$II. Institut für Theoretische Physik, Universität Hamburg, Luruper Chaussee
149, 22761, Hamburg, Germany}\\
{\it $^{\, f}$Institute for Theoretical Physics, ELTE Eötvös Loránd University, Pázmány Péter sétány
1/A, 1117, Budapest, Hungary}\\
{\it $^{\, g}$Physik-Institut, Universität Zürich, 8057 Zürich, Switzerland}\\
{\it $^{\, h}$Dipartimento di Fisica Ettore Pancini, Universit`a di Napoli Federico II and INFN - Sezione di
Napoli, Complesso Universitario di Monte Sant’Angelo Ed. 6, Via Cintia, 80126 Napoli, Italy}
\vspace{1.4cm}

{\large \bf Abstract}
\vspace{-0.2cm}
\end{center}
We give an update on the status of \nnlocal, a parton-level Monte Carlo program implementing the extension of the completely local subtraction scheme \colorful to color-singlet production in hadron-hadron collisions. The construction of the counterterms in our scheme is generic, being based on the standard IR factorization formulae of QCD. Furthermore, the integration of the counterterms over the phase space of unresolved emissions is performed fully analytically, allowing for good control of the numerical stability of our predictions. We validate our method by computing NNLO corrections to fully differential cross sections for the LHC.
\end{titlepage}
\clearpage

%%%
%%% TOC here
%%%

%\tableofcontents

%\renewcommand{\thefootnote}{\fnsymbol{footnote}}

%%%%%%%%%%%%%%%%%%%%%%%%%%%%%%%%%%%%%%%%%%%%%%%%%%%%%%%%%%%%
%%%%%%%%%%%%%%%%%%%%%%%%%%%%%%%%%%%%%%%%%%%%%%%%%%%%%%%%%%%%

%%%
%%% Notation
%%%

\section{Introduction}
\label{sec:Introduction}
The Standard Model (SM) successfully explains particle interactions over a wide range of energies. Despite this, it is known to be incomplete. Open questions, such as the nature of dark matter and dark energy, the imbalance between matter and antimatter, and the origin of neutrino masses, point to physics \textit{beyond} the SM. So far, there is no conclusive evidence for new particles or interactions. Instead, new physics may reveal itself indirectly through small deviations between theoretical predictions and experimental results. As such, precision is key. On the theory side, this means calculating higher-order corrections to perturbative cross sections, which is challenging due to divergences that produce infinities at intermediate stages of the computation. Ultraviolet (UV) divergences in virtual corrections are handled through renormalization, while infrared (IR) divergences, arising in both real and virtual contributions, cancel in physical observables according to the KLN theorem \cite{Nakanishi:1958gt,Kinoshita:1962ur,Lee:1964is} combined with mass factorization. So, if we could perform the calculations fully analytically, there would be no real issue. Unfortunately, practical predictions typically require fully differential results, making analytic solutions infeasible. Therefore, IR singularities must be treated carefully to allow for numerical phase space integration. At next-to-leading order, reliable methods and automation have largely solved this problem, driving the NLO revolution of the 2010s \cite{Frixione:1995ms,Catani:1996vz}. Perhaps the rise of machine learning and AI will lead to the next revolution in the field, see e.g. \cite{Heimel:2022wyj,DeCrescenzo:2026tsp} for steps in this direction. 

The computation of next-to-next-to-leading-order (NNLO) corrections is a
highly active field of research with several approaches available in the
literature. These are typically based on two fundamentally different
strategies, namely phase space slicing \cite{Fabricius:1981sx} and local subtraction
\cite{Ellis:1980wv}. The main idea of the former is to introduce a small cut-off parameter to regulate the divergent phase space integrals. Examples include $q_T$-subtraction \cite{Catani:2007vq} and $N$-jettiness subtraction \cite{Gaunt:2015pea}, both of which are available through public tools such as {\tt MCFM} \cite{Boughezal:2016wmq} and {\tt MATRIX} \cite{Grazzini:2017mhc}. For methods based on local subtraction, one aims to construct approximate cross sections that match the
point-wise singularity structure of the partonic cross sections. As such, subtraction leads to inherently finite phase space integrals. Next, the subtraction terms are integrated over the unresolved phase space and added back, taking care of explicit $\varepsilon$ poles coming from virtual corrections. Many examples of subtraction schemes are available in the literature \cite{Gehrmann-DeRidder:2005btv,Caola:2017dug,Czakon:2014oma,Magnea:2018hab,Cacciari:2015jma,Herzog:2018ily,Somogyi:2005xz}, which differ either in the way the subtraction terms are defined or how overlapping limits are treated. Until recently, there was a clear lack of public implementations of such schemes. However, very recently, the situation started improving, with the public codes {\tt NNLOJET} \cite{NNLOJET:2025rno} and {\tt history} \cite{Klein:2026tlj} becoming available. The former offers an implementation of antenna subtraction, while the latter is based on nested soft-collinear subtraction\footnote{Note, however, that the original authors of the nested soft-collinear method are not involved in the production of {\tt history}, and are in fact working on their own program based on improvements presented during this conference.}. In this proceeding, we focus on another publicly available Monte Carlo program, namely \nnlocal \cite{DelDuca:2024ovc}, which implements the extension of the \colorful subtraction scheme to hadronic processes.
The remainder of this text is organized as follows. In the next section, we review the regularization of IR divergences in the \colorful subtraction scheme. Sec.~\ref{sec:NNLOCALpub} then presents a brief recap of the features of the public version of \nnlocal, while in sec.~\ref{sec:NNLOCALdev} we highlight some ongoing developments. Section \ref{sec:summary} provides a brief summary and outlook on what will come next.

\section{\colorful for hadron collisions}
Consider a hadron-hadron collision with the production of a colorless state
$X$ in association with $m$ jets. This could, for example, represent Higgs + jet production at the LHC. Because of QCD factorization, the cross section of such a process can be written as
\begin{equation}
        \hat{\sigma}(p_A,p_B) = \sum_{a,b}
        \int_0^1 \rd x_a \, f_{a/A}(x_a,\mu_F^2) \int_0^1 \rd x_b \, f_{b/B}(x_b,\mu_F^2)\,
    \sigma_{ab}(p_a,p_b;\mu_F^2)\,.
\end{equation}
Put in words, the hadronic cross section corresponds to a convolution of the non-perturbative parton distribution functions (PDFs) with the perturbative partonic cross section,
\begin{equation}
    \sigma_{ab}(p_a,p_b;\mu_F^2) =
    \sum_{k=0}^{\infty}\sigma^{\text{N$^k$LO}}_{ab}(p_a,p_b;\mu_R^2,\mu_F^2)\,.
\end{equation}
The partonic momenta are defined in terms of the hadronic ones as $p_{a/b}=x_{a/b}p_{A/B}$. In the following, we will suppress the arguments of the partonic cross section.
At NNLO accuracy, we can write
\begin{equation}
\label{eq:xsNNLO}
    \begin{split}
        \tsigk{NN}{}{LO}{}_{ab} =\, 
    \int_{m+2} \dsigk{RR}{}{}{}_{ab} J_{m+2}
    +\int_{m+1}\left(\dsigk{RV}{}{}{}_{ab} + \dsigk{C}{1}{}{}_{ab}\right) J_{m+1}
    +\int_{m}\left(\dsigk{}{}{VV}{}_{ab} + \dsigk{C}{2}{}{}_{ab}\right) J_{m}\,.
    \end{split}
\end{equation}
Here, $\dsigk{RR}{}{}{}_{ab}$, $\dsigk{RV}{}{}{}_{ab}$ and $\dsigk{VV}{}{}{}_{ab}$ represent the double-real, real-virtual and double-virtual contributions. The collinear remnants, coming from the renormalization of the PDFs, are denoted by $\dsigk{C}{1}{}{}_{ab}$ and $\dsigk{C}{2}{}{}_{ab}$. For an infrared- and collinear-safe observable $J_n$, the cross section in eq.~(\ref{eq:xsNNLO}) is finite, but each  phase space integral is separately IR divergent. In the \colorful scheme, the IR regularization is implemented as follows
\begin{equation}
\label{eq:XSreg}
    \begin{split}
&\tsig{NNLO}_{ab,\text{reg}} =\, 
    \int_{m+2}\left[
    \dsig{RR}_{ab} J_{m+2} - \dsiga{RR}{1}_{ab} J_{m+1} - \dsiga{RR}{2}_{ab} J_{m} + \dsiga{RR}{12}_{ab} J_{m} 
    \right]
\\ &\quad+
    \int_{m+1}\left\{
    \left[\dsig{RV}_{ab} + \dsigk{C}{1}{}{}_{ab} + \int_1 \dsiga{RR}{1}_{ab}\right] J_{m+1}
    -\left[\dsiga{RV}{1}_{ab} + \dsigk{C}{1,}{A}{1}_{ab} + \left(\int_1 \dsiga{RR}{1}_{ab}\right)^{\!{\rm A}_1}\right] J_{m}
    \right\}
\\ &\quad+
    \int_{m} \Bigg\{
    \dsig{VV}_{ab} + \dsigk{C}{2}{}{}_{ab} + \int_2\left[\dsiga{RR}{2}_{ab} - \dsiga{RR}{12}_{ab}\right] + \int_1\left[\dsiga{RV}{1}_{ab} + \dsigk{C}{1,}{A}{1}_{ab}\right]\\&\quad + \int_1\left(\int_1 \dsiga{RR}{1}_{ab}\right)^{\!{\rm A}_1}
    \Bigg\} J_{m}\,.
\end{split}
\end{equation}
Each approximate cross section is specifically designed to take care of some divergence as one or two partons become unresolved. In particular, approximate cross sections denoted with $A_1$ regulate single unresolved emissions while those coming with $A_2$ regulate double unresolved ones. The one exception is $\dsiga{RR}{12}_{ab}$, which accounts for the fact that the $A_1$ and $A_2$ approximations overlap in phase space, and so its inclusion avoids over-subtraction. Specifically, it is constructed to take care of single unresolved limits of $\dsiga{RR}{2}_{ab}$ and double unresolved ones of $\dsiga{RR}{1}_{ab}$. 

By construction, the three phase space integrals in eq.~(\ref{eq:XSreg}) are IR finite and hence can be evaluated using standard numeric tools. Note that explicit poles in $\varepsilon$, coming from loop integrals, are cancelled by the $\varepsilon$ poles coming from the integrated subtraction terms. As was explained in detail in \cite{DelDuca:2025yph}, each subtraction term is systematically derived from the known infrared limit formulae of squared QCD matrix elements. The momentum fractions, eikonal factors etc. that appear in such formulae are carefully extended beyond the strict IR limits in order for our subtraction terms to be properly defined over the full phase space. Moreover, the issue of overlapping limits is treated using the inclusion-exclusion principle (i.e., subtract all limits, add back the double overlaps etc.).

The integrals of the subtraction terms over the momenta of the unresolved emissions are computed fully analytically. This way, analytic pole cancellation can be checked explicitly, ensuring correctness of the subtraction scheme. For now, this was done with color-singlet production ($m = 0$) in mind. We want to emphasize however that most of our results are in fact completely generic, and can be carried over to processes involving jet production \textit{without} modification.\footnote{The few exceptions to this are subtraction terms involving eikonal factors with hard partons. As in color-singlet production we only have two hard partons, which moreover are back-to-back, the kinematics simplify considerably. Beyond color-singlet, more hard partons appear with more general kinematics, and hence the corresponding integrals will have to be recomputed.}

All-in-all, there are 256 basic integrals to compute. In practice, it turns out to be beneficial to employ different strategies for different sets of integrals. A subset of the counterterms is integrated using integration-by-parts reduction \cite{Chetyrkin:1981qh}, in combination with the method of differential equations \cite{Henn:2013pwa}. For example, the $A_2$ counterterms are all integrated in this fashion \cite{Guadagni:2026bby,DelDuca:2026tdg}. Another approach used to compute the unresolved phase space integrals is by direct integration. Here, we exploit the fact that, due to the precise definitions of momentum mappings in our scheme, the real-emission phase space can be written as a convolution between the mapped and unresolved phase spaces. This allows us to write down a \textit{parametric representation} of the integrals, which can be evaluated using standard methods in terms of generalized polylogarithms \cite{Goncharov:1998kja}. For example, all $A_{12}$ subtraction terms are integrated like this \cite{VanThurenhout:2024hmd,Fekeshazy:2025ktp}. A surprising bottleneck in these computations was the need to perform univariate partial fraction decompositions of large multivariate expressions. Publicly available tools, such as the standard {\tt Apart} command in {\tt Mathematica}, were not up to the task as time and memory consumption exploded. We circumvented this issue by developing a new routine, called {\tt LinApart}, which is based on a closed formula for the decomposition following directly from the residue theorem \cite{Chargeishvili:2024nut}. Our routine, whose implementations in {\tt Mathematica} and {\tt C} are publicly available at \url{https://github.com/fekeshazy/LinApart}, leads to significant improvements (typically several orders of magnitude) in both time and memory usage.\footnote{Our original routine, developed for the integration of the subtraction terms, only treated denominators which were fully factorized in the decomposition variable. However, recently, an updated version was released that can deal with \textit{arbitrary} denominators \cite{Fekeshazy:2025mlh}.} While we will not go into detail here on the exact functional form of the integrated counterterms, we want to highlight one important difference compared to processes with colorless initial states. For the latter, the integrated subtraction terms are simply \textit{functions} of momentum fractions of initial-state momenta. This is no longer true for hadron collisions. In particular, the integrals are now \textit{distributions} acting on the PDFs, meaning they can be written in terms of (double) plus distributions and Dirac delta distributions.

\section{The public version of {\tt NNLOCAL}}
\label{sec:NNLOCALpub}
The definitions of the various subtraction terms, together with their integrated versions, have been implemented in the parton-level Monte Carlo program \nnlocal. The code, which is written in {\tt Fortran77} and whose architecture is based on {\tt MCFM-4.0} \cite{Campbell:2010ff}, is publicly available at \url{https://github.com/nnlocal/nnlocal.git}. Its only external dependency is {\tt LHAPDF} \cite{Buckley:2014ana} for the evaluation of the PDFs. For now, the public version serves as a proof-of-concept of our method. In particular, the only process currently implemented is gluon-fusion Higgs production in HEFT without light quarks (i.e., $n_f=0$). We emphasize, however, that this is not a limitation on our subtraction scheme. In fact, the all-gluon subprocess has the most involved IR structure that can appear at NNLO accuracy, as it gets contributions from \textit{all} IR limits. As such, concentrating on this particular process allows us to highlight the functionality of our method without having to worry about details which are irrelevant for the IR subtraction (such as, e.g., flavor combinatorics when $n_f\neq 0$).

\begin{figure}[H]
    \includegraphics[width=0.3\linewidth]{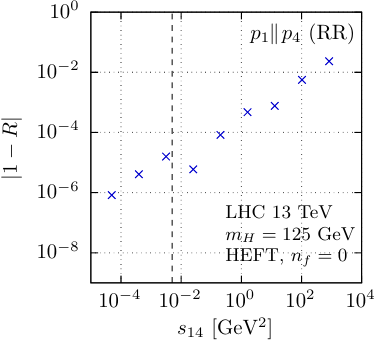}
    \hspace{1.5em}
    \includegraphics[width=0.3\linewidth]{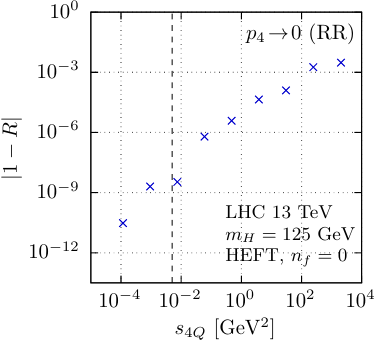}
    \hspace{1.5em}
    \includegraphics[width=0.3\linewidth]{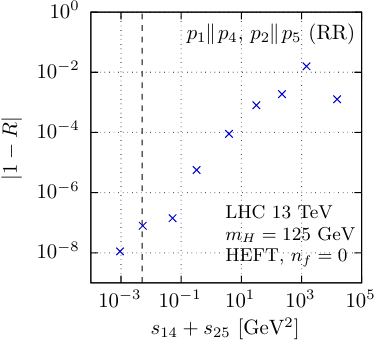}
    \caption{Cancellation of kinematic singularities in various IR limits. From left to right, we show a single-collinear, single-soft and double-collinear limit. The horizontal axis shows an appropriate kinematic variable whose smallness characterizes the IR limit of interest. On the vertical axis, we plot $|1-R|$, with $R$ the ratio of the sum of subtraction terms and the double-real contribution to the NNLO cross section. As expected, $R$ converges to one as we go deeper into the limit. The dashed line represents a technical cutoff, which is necessary due to finite-precision arithmetic. The independence of physical observables on the value of this cutoff was carefully checked.}
\label{fig:1}
\end{figure}

\begin{figure}[H]
   \centering
   \includegraphics[width=0.4\linewidth]{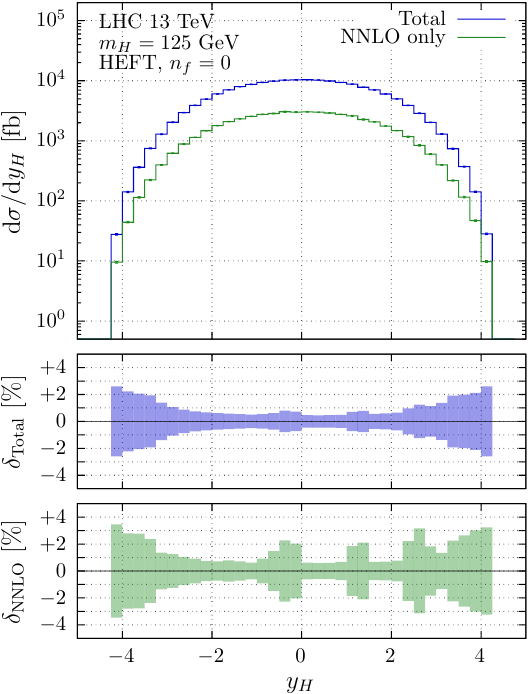}
   \hspace{1.5em}
   \includegraphics[width=0.4\linewidth]{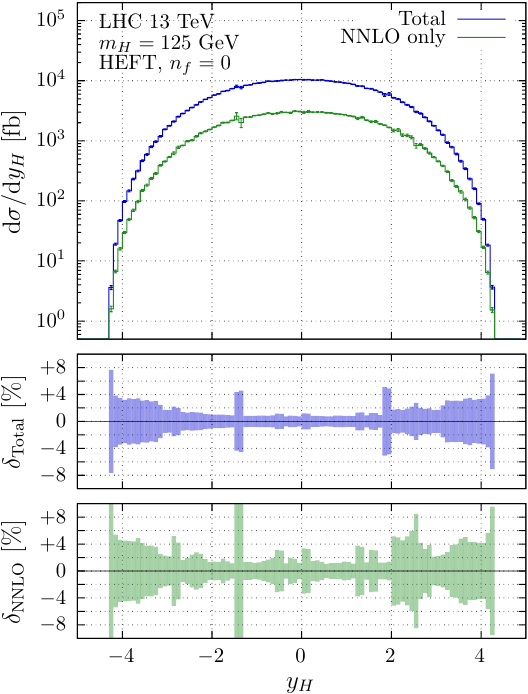}
   \caption{Rapidity distributions of the Higgs boson with bin widths $\Delta y = 0.25$ (left) and $\Delta y = 0.1$ (right). The errors represent the estimated Monte Carlo uncertainties. The runtime to produce these plots was about 1 hour and 15 minutes on a MacBook Pro M2 with 8 CPUs.}
\label{fig:2}
\end{figure}

A detailed account of the features of our code was provided in \cite{DelDuca:2024ovc}, and here we limit ourselves to a brief summary.
\begin{enumerate}
    \item Routines are in place to allow the user to directly validate the cancellation of $\eps$-poles.
    \item Similarly, the cancellation of kinematic singularities can be checked explicitly, as illustrated in figure \ref{fig:1}.
    \item In principle, one can compute any infrared- and collinear-safe observable, and both inclusive and exclusive predictions can be produced. Examples are given in table \ref{tab:1} and figure \ref{fig:2}.
    \item Explicit support for running in parallel mode is provided through shell scripts.
\end{enumerate}

\begin{table}[H]
\setlength{\tabcolsep}{12pt}
\renewcommand{\arraystretch}{1.5}
    \centering
    \begin{tabular}{|c|c|c|}
    \hline%\hline
         $m_H$ [GeV] &  {\tt n3loxs} (gg) &  {\tt
         NNLOCAL} (gg) \\
    \hline\hline
    100 & $65.72$ pb & $65.74 (4)$ pb \\
    \hline
         125 & $42.94$ pb & $42.94(2)$ pb\\
    \hline
         250 & $9.730$ pb & $9.733(5)$ pb \\
    \hline
         500 & $1.626$ pb & $1.626(1)$ pb \\
    \hline
         1000 & $173.7$ fb & $173.7(1)$ fb \\
    \hline
    2000 & $8.794$ fb & $8.790 (5)$ fb \\
    \hline
    \end{tabular}
    \caption{Predictions for the fully inclusive NNLO cross section for gluon-fusion Higgs production in $n_f=0$ HEFT at the 13 TeV LHC for various values of the Higgs mass. The scales are set to $\mu_R=\mu_F=m_H$. To validate our results, we compare with {\tt n3loxs} \cite{Baglio:2022wzu}. The quoted errors correspond to Monte Carlo uncertainties. The runtime per mass value is about 20 minutes on a MacBook Pro M2 with 8 CPUs.}
\label{tab:1}
\end{table}

\section{Current developments}
\label{sec:NNLOCALdev}
Next, we discuss some recent updates which, for now, are not yet public. We have now implemented \textit{all} partonic subprocesses for $pp\to H$ at NNLO accuracy. As illustration, we present the contribution of each separate subprocess to the double-real correction in table \ref{tab:2} and to the real-virtual one in table \ref{tab:3}. Of course, these numbers themselves have no physical meaning. However, the fact that we find convergence to numerically stable results provides a strong check on our framework and implementation. We also want to emphasize that the errors, originating from the Monte Carlo integration, are nicely under control. For example, for the $gg$ subprocess, we see from table \ref{tab:1} that the total cross section is of the order of $10^4$ fb, while from tables \ref{tab:2} and \ref{tab:3} we see that the corresponding error is only a few fb.  
\begin{table}[h]
\centering
\begin{minipage}{0.45\textwidth}
\centering
\begin{tabular}{|c|c|c|}
    \hline%\hline
         Subprocess & $\sigma^{\text{RR,reg}}$ (fb)  & \% \\
    \hline\hline
         $gg$ & $180.7\pm 3.6$ & $40.06$ \\
    \hline
         $gq$ & $166.7\pm 3.4$ & $36.96$ \\
    \hline
         $g\Bar{q}$ & $55.87\pm 0.48$ & $12.39$ \\
    \hline
         $qq$ & $27.49\pm 0.01$ & $6.09$ \\
    \hline
         $q\Bar{q}$ & $18.17\pm 0.01$ & $4.03$ \\
    \hline
    $\Bar{q}\Bar{q}$ & $2.121\pm 0.001$ & $0.47$ \\
    \hline
         $\Sigma$ & $451.1\pm 7.5$ & $100$ \\
    \hline
    \end{tabular}
\caption{The double-real contributions to the total Higgs boson production cross
section at the 13 TeV LHC for the various partonic subprocesses.}
\label{tab:2}
\end{minipage}
\hfill
\begin{minipage}{0.45\textwidth}
\centering
 \begin{tabular}{|c|c|c|}
    \hline%\hline
         Subprocess & $\sigma^{\text{RV,reg}}$ (fb)  & \% \\
    \hline\hline
         $gg$ & $-1393\pm 1 $ & $62.06$ \\
    \hline
         $gq$ & $-610.4\pm 0.6 $ & $27.19$ \\
    \hline
         $g\Bar{q}$ & $-240.1\pm 0.2 $ & $10.69$ \\
    \hline
         $qq$ & $-14.78\pm 0.02$ & $0.66$ \\
    \hline
         $q\Bar{q}$ & $14.45\pm 0.02 $ & $-0.64$ \\
    \hline
    $\Bar{q}\Bar{q}$ & $-0.9630\pm 0.0014$ & $0.04$ \\
    \hline
         $\Sigma$ & $-2245\pm 2$ & $100$ \\
    \hline
    \end{tabular}
\caption{The real-virtual contributions to the total Higgs boson production cross
section at the 13 TeV LHC for the various partonic subprocesses.}
    \label{tab:3}
\end{minipage}
\end{table}
We do not show a separate table for the VV line. However, analytic pole cancellation for all partonic subprocesses has been explicitly checked. As for the implementation of the corresponding expression into \nnlocal, we are currently working on optimizing its evaluation. To understand why this is necessary, remember that the VV line is where the majority of integrated counterterms are collected. The latter are large expressions, depending on the two convolution variables that enter the PDFs. I.e., symbolically, we can write a generic integrated subtraction term as
\begin{equation}
\label{eq:ICT}
    \begin{split}
        \left[ \CT\right]\sim\,&\,\int_{0}^{1}\rd x_a\,\rd x_b\,\rd\sigma_{a b}(x_a p_A,x_b p_B)\int_{0}^{1}\rd \eta_a\,\rd \eta_b\,{[\CT(\eta_a,\eta_b;\eps)]} \frac{f_{a/A}(x_a/\eta_a)}{\eta_a}\frac{f_{b/B}(x_b/\eta_b)}{\eta_b}\,.
    \end{split}
\end{equation}
Looking at eq.~(\ref{eq:ICT}), one might be worried about singularities as $\eta_a$ and/or $\eta_b$ approach zero or one. Indeed, there are singularities when either (or both) of the variables approach one. These are treated explicitly by way of subtraction. I.e., we compute the asymptotic behaviour in all relevant limits using
expansion by regions \cite{Beneke:1997zp}, subtract the limit formulae and add back their integrated versions. It turns out that, for physical processes, the convolution variables cannot become zero (in particular, we have kinematic restrictions of the form $\eta_a \eta_b > M_X^2/s_{ab}$). Nevertheless, the numeric evaluation becomes unstable as $\eta_a$ and/or $\eta_b$ becomes small. This of course is a well-known issue. Currently, we circumvent this by dynamically switching to quadrupole precision when $\eta_{a,b}<5\times 10^{-3}$. In practice, such cases are rare (for example, for physical Higgs production, we need to go to quad precision for about 6\% of the evaluations). Nevertheless, it causes the code to slow down significantly. As such, we are working to remove the requirement to go to quad precision altogether. For this, we can of course exploit the fact that our integrated counterterms are fully analytic, and we are currently investigating the best way forward.

Finally, let us briefly turn to some updates in the code itself. We now have routines in place to study both scale variations (e.g. by 7-point variation) and PDF errors automatically. Furthermore, we are currently working on enhancing the flexibility of the code by completely separating the implementation of the IR subtraction. This way, the code will be easily applicable to any color-singlet process once the relevant matrix elements are provided.

\section{Summary and outlook}
\label{sec:summary}
We have presented an update on the parton-level Monte Carlo program \nnlocal, which implements the extension of the \colorful subtraction scheme to hadron-hadron collisions. In particular, while the public version of the code is still a proof-of-concept, we are making progress in turning it into a useful tool. At the moment, we are aiming at enhanced efficiency (by removing the need to go to quadrupole precision) and enhanced flexibility (by separating the IR subtraction from the rest of the code). Once this is complete, we can start building a library for color-singlet processes.

Finally, the extension of our method to both jet production and higher orders in perturbation theory is feasible. In particular, we want to stress that most of our integrated subtraction terms, while computed for color-singlet processes, are in fact completely generic. This means that they can be used for processes with jets in the final state without modification. As such, to extend our methodology to jet processes, we only need to account for (a) subtraction terms which are explicitly missing for color-singlet production and (b) integrated counterterms whose definitions change explicitly once final-state jets are included. These considerations are left for future studies.

\subsection*{Acknowledgements}
This work has been supported by grant K143451 of the National Research, Development and Innovation Fund in Hungary, the Bolyai Fellowship program of the Hungarian Academy of Sciences and by the Deutsche Forschungsgemeinschaft (DFG, German Research Foundation) under Germany’s Excellence Strategy – Cluster of Excellence “Color meets Flavor”, EXC 3107 – Project-ID 533766364. The work of C.D. was funded by the European Union (ERC Consolidator Grant LoCoMotive 101043686). Views and opinions expressed are however those of the author(s) only and do not necessarily reflect those of the European Union or the European Research Council. Neither the European Union nor the granting authority can be held responsible for them. The work of L.F.\ was supported by the German Academic Exchange Service (DAAD) through its Bi-Nationally Supervised Scholarship program.

{\scriptsize
\bibliographystyle{JHEP}
\bibliography{nnlocal}
}

%%%
%%% End document
%%%

\end{document}